\documentclass[sigconf, authorversion]{acmart} %

\AtBeginDocument{%
  }

\copyrightyear{2022} 
\acmYear{2022} 
\setcopyright{acmlicensed}\acmConference[ACSAC '22]{Annual Computer Security Applications Conference}{December 5--9, 2022}{Austin, TX, USA}
\acmBooktitle{Annual Computer Security Applications Conference (ACSAC '22), December 5--9, 2022, Austin, TX, USA}
\acmPrice{15.00}
\acmDOI{10.1145/3564625.3564628}
\acmISBN{978-1-4503-9759-9/22/12}

\usepackage{algorithmic}
\usepackage{textcomp}
\usepackage{dsfont}
\usepackage{tabularx}
\newcolumntype{Y}{>{\centering\arraybackslash}X} %
\usepackage{xcolor}
\usepackage{enumitem}
\usepackage{xspace}
\usepackage{hyperref} %
\usepackage[binary-units]{siunitx}
\ifdefined\unit\else
\ifdefined\NewCommandCopy
\NewCommandCopy\unit\si
\else
\NewDocumentCommand\unit{O{}m}{\si[#1]{#2}}
\fi
\fi
\newlist{questions}{enumerate}{1}
\setlist[questions,1]{label=\textbf{RQ\arabic*.},ref=\textbf{RQ\arabic*}}
\usepackage[section,numberedsection=autolabel,nogroupskip,nonumberlist]{glossaries}
\makeglossaries
\newacronym{poi}{POI}{Point of Interest}
\newacronym{sdd}{SDD}{Sampling Distance and Direction}
\newacronym{gan}{GAN}{Generate Adversarial Network}
\newacronym{relu}{ReLU}{Rectified Linear Activation}
\newacronym{tanh}{tanh}{Hyperbolic Tangent}
\newacronym{mlp}{MLP}{Multi-Layer Perceptron}
\newacronym{tul}{TUL}{Trajectory User Linking}
\newacronym{TRA}{TRA}{Trajectory Reconstruction Attack}
\newacronym{RAoPT}{RAoPT}{Reconstruction Attack on Protected Trajectories}
\newacronym{lstm}{LSTM}{Long Short-Term Memory}
\newacronym{rnn}{RNN}{Recurrent Neural Network}
\newacronym{mse}{MSE}{Mean Squared Error}
\newacronym{mae}{MAE}{Mean Absolute Error}
\newglossaryentry{OR-Distance}
{
    name=OR-Distance,
	description={Distance between \emph{original} and \emph{reconstructed} traj.}
}
\newglossaryentry{OP-Distance}
{
    name=OP-Distance,
    description={Distance between  \emph{original} and \emph{protected} trajectory}
}
\makeatletter\let\myfsize\f@size\makeatother
\newcommand{\subheading}[1]{\noindent{\textbf{#1}}}

\begin{document}

\title{Reconstruction Attack on Differential Private Trajectory Protection Mechanisms}

\author{Erik Buchholz}
\orcid{0000-0001-9962-5665}
\affiliation{%
  \institution{University of New South Wales}
  \institution{Data61, Cyber Security CRC}
  \city{Sydney}
  \state{NSW}
  \country{Australia}
}
\email{e.buchholz@unsw.edu.au}

\author{Alsharif Abuadbba}
\orcid{0000-0001-9695-7947}
\affiliation{%
  \institution{CSIRO's Data61}
  \institution{Cyber Security CRC}
  \city{Sydney}
  \state{NSW}
  \country{Australia}
}
\email{sharif.abuadbba@data61.csiro.au}

\author{Shuo Wang}
\orcid{0000-0001-8938-2364}
\affiliation{%
  \institution{CSIRO's Data61}
  \institution{Cyber Security CRC}
  \city{Sydney}
  \state{NSW}
  \country{Australia}
}
\email{shuo.wang@data61.csiro.au}

\author{Surya Nepal}
\orcid{0000-0002-3289-6599}
\affiliation{%
 \institution{CSIRO's Data61}
 \institution{Cyber Security CRC}
 \city{Sydney}
 \state{NSW}
 \country{Australia}
 }
 \email{surya.nepal@data61.csiro.au}

\author{Salil S. Kanhere}
\orcid{0000-0002-1835-3475}
\affiliation{%
  \institution{University of New South Wales}
  \institution{Cyber Security CRC}
  \city{Sydney}
  \state{NSW}
  \country{Australia}
}
\email{salil.kanhere@unsw.edu.au}

\begin{abstract}
Location trajectories collected by smartphones and other devices represent a valuable data source for applications such as location-based services.
Likewise, trajectories have the potential to reveal sensitive information about individuals, e.g., religious beliefs or sexual orientations. 
Accordingly, trajectory datasets require appropriate sanitization. 
Due to their strong theoretical privacy guarantees, differential private publication mechanisms receive much attention. 
However, the large amount of noise required to achieve differential privacy yields structural differences, e.g., ship trajectories passing over land.
We propose a deep learning-based \emph{\gls{RAoPT}}, that leverages the mentioned differences to partly reconstruct the original trajectory from a differential private release.
The evaluation shows that our \gls{RAoPT} model can reduce the Euclidean and Hausdorff distances between the released and original trajectories by over \SI{68}{\%} %
on two real-world datasets under protection with $\varepsilon \leq 1$.
In this setting, the attack increases the average Jaccard index of the trajectories' convex hulls, representing a user's activity space, by over \SI{180}{\%}.
Trained on the GeoLife dataset, the model still reduces the Euclidean and Hausdorff distances by over \SI{60}{\%} for T-Drive trajectories protected with a state-of-the-art mechanism ($\varepsilon = 0.1$). %
This work highlights shortcomings of current trajectory publication mechanisms, and thus motivates further research on privacy-preserving publication schemes.
\end{abstract}



\maketitle

\section{Introduction}\label{sec:intro}

Due to the omnipresence of smartphones and wearables in our daily lives, a large amount of personal location data is collected every day.
The sequence of locations visited by an individual represents a trajectory. 
This data is valuable for many services such as research~\cite{Barbosa2018}, market analysis~\cite{Xu2011}, navigation~\cite{WazeMobile, Google}, social gaming~\cite{Niantic}, and most recently for contact tracing in the context of the COVID-19 pandemic~\cite{Pokharel2021,Kleinman2020,Park2020}. 
However, significant privacy concerns are associated with the release of location information.
The location trajectory of an individual may reveal sensitive information, such as religious, political, or sexual beliefs~\cite{Primault2019, Abul2008}.
For instance, someone with access to the trajectories of a taxi fleet could determine which drivers are practising Muslims based on the correlation of their breaks and mandatory prayer times~\cite{Franceschi-Bicchierai}.
Moreover, De Montjoye et al.\ showed that only four spatio-temporal points suffice to uniquely identify \SI{95}{\%} of individuals~\cite{DeMontjoye2013}.
These examples illustrate the need of trajectories for appropriate protection before being released. 

To hide the exact route of individual trajectories with the goal to prevent pirate attacks \cite{Jiang2013}, stalking \cite{Gomes2018} or other security threats, multiple approaches that extend $k$-anonymity \cite{Abul2008, Xin2017, Gomes2018, Nergiz2008, Abul2010, Shaham2021} and differential privacy
\cite{Chen2011,Li2017,Liu2021a,Hua2015,Cao2021,Jiang2013}
to the trajectory domain have been proposed~\cite{Primault2019}.
However, existing approaches either significantly reduce the utility of the released data or provide limited privacy~\cite{Primault2015, Primault2019}. 
$K$-anonymity based approaches are susceptible to attacks utilising background knowledge, and cannot provide strong privacy guarantees~\cite{Chen2020, Jiang2022, Chen2011, Ma2021, Gomes2018}.
Therefore, recent research focused on publication mechanisms achieving differential privacy.
Due to the high information content of location data, these approaches significantly degrade data utility to achieve privacy protection~\cite{Rao2020, Qu2020, Ma2021} because there is an inherent trade-off between privacy and utility. 
The random distortion added by differential private protection mechanisms yields unrealistic trajectories that provide limited utility and can easily be recognised~\cite{Qu2020} because they do not take geographical constraints into consideration~\cite{Naghizade2020}.
For instance, protected trajectories of cars do not follow roads or ship trajectories pass over land. %
To highlight the risk posed by current publication mechanisms, we address the research question:
\begin{questions}
   \item Can an adversary (partly) reconstruct trajectories from a differential private trajectory release? \label{rq1}
\end{questions}
We find that the differential private mechanisms, such as the \gls{sdd} approach \cite{Jiang2013}, yield trajectories that are structurally distinguishable from unperturbed trajectories.
By exploiting these characteristics, we propose a novel \gls{lstm}-based \glsfirst{RAoPT} to address the defined research question \ref{rq1}.
The \gls{RAoPT} model receives trajectories protected with a differential private publication mechanism as input and outputs reconstructed trajectories which are closer to the original trajectories.

Our attack is evaluated on two real-world GPS datasets, the T-Drive \cite{t-drive}, and the GeoLife \cite{Geolife1} dataset.
We evaluate the reduction of the Euclidean and Hausdorff distances between the recovered and original trajectories compared to the distances between the protected and original trajectories.
These metrics are commonly used to measure the distance between trajectories \cite{Jiang2013, Shao2020, Rao2020,Shao2020,Ma2021,Hua2015,Li2017}.
We also measure the increase of the Jaccard index of the trajectories' convex hulls before and after reconstruction, as the convex hull can represent a trajectory's activity space, i.e., the area in which a user is active \cite{Lee2016}.
A small physical distance to a victim represents a security threat in various settings, e.g., stalkers can follow or intercept a victim \cite{Gomes2018}, or pirates can plan attacks \cite{Jiang2013}.

For an adversary with knowledge about the used protection method and access to ground truth trajectories for the training, our \gls{RAoPT} model can reduce both distances even for privacy settings with very high privacy guarantees ($\varepsilon \leq 0.1$) by over \SI{98}{\%} %
in case of Laplace noise-based protection and by \SI{68}{\%} to \SI{84}{\%} %
considering a state-of-the-art protection mechanism.
In these settings, the Jaccard index is increased by at least \SI{180}{\%}.
In the realistic scenario that the adversary knows about the protection method (\gls{sdd} with  $\varepsilon = 0.1$) but no ground truth for training is accessible, the Euclidean distance can still be reduced by approx. \num{40} to \SI{61}{\%}, %
and the Hausdorff distance by approx. \SI{62}{\%} to \SI{67}{\%} when transferring from one dataset to the other.
Even an adversary without background knowledge can achieve reductions of over \SI{60}{\%} in some settings.

\subheading{Contributions.} Our main contributions are as follows:
\begin{itemize}
    \item We propose the first \gls{lstm}-based reconstruction attack on differential private protected trajectories. 
    \item The \gls{RAoPT} model can reduce the Euclidean and Hausdorff distances by over \SI{68}{\%}%
    on T-Drive trajectories protected with the Laplace mechanism or a state-of-the-art protection mechanism and $\varepsilon \leq 1$, decreasing the provided level of privacy. 
    \item We show the real-world applicability of the attack via two datasets, namely T-Drive~\cite{t-drive} and GeoLife~\cite{Geolife1}, with different granularities and transport modes.
    \item We open-source our \gls{RAoPT} model\footnote{Our source code is available at: \url{https://github.com/erik-buchholz/RAoPT}}
        including the considered protection mechanisms and pre-processing scripts.
\end{itemize}
 
This article is organised as follows.
We introduce the required background knowledge in Section~\ref{sec:background} and define our problem statement and threat model in Sections~\ref{sec:problem-statement} and~\ref{sec:threat-model}, respectively. 
In Section~\ref{sec:related-work}, we discuss related work addressing attacks on protection mechanisms, and approaches utilising deep learning for trajectory protection.
Then, we introduce our proposed \gls{RAoPT} model in Section~\ref{sec:attack}.
In Section~\ref{sec:eval}, we provide implementation details and show the results of our evaluations.
Subsequently, we discuss our findings in Section~\ref{sec:discussion}.
Finally, we conclude this paper in Section~\ref{sec:conclusion}.

\section{Preliminaries}\label{sec:background}
In Section~\ref{sec:trajectory},  we first define the term \emph{location trajectory} and provide some general knowledge on trajectory protection mechanisms.
The \gls{sdd} mechanism, which is used for the evaluation of our attack, is explained in Section~\ref{sec:sdd}.
In Section~\ref{sec:problem-statement}, we state our problem statement, and in Section~\ref{sec:threat-model}, we present our threat model.
We refer readers not familiar with differential privacy to Appendix~\ref{sec:dp}.

\subsection{Location Trajectory Protection}\label{sec:trajectory}

A location trajectory $T$ consists of a sequence of locations $T = (t_1, \dots, t_n)$. 
In the most basic case, each location consists of two values $t_i = (x_i, y_i)$ which can either represent the location within a reference (coordinate) system or \emph{latitude} and \emph{longitude}.
For exact localisation, the \emph{altitude} or \emph{elevation} can be added to latitude and longitude as a third coordinate.
Many trajectory datasets record a \emph{timestamp} for each location \cite{t-drive, Geolife1}. 
Lastly, trajectories can be enhanced by semantic information \cite{Tu2019}, such as \glspl{poi}, i.e., the knowledge of whether a location within a trajectory represents a restaurant, shop, or gym.
While such additional information can increase the utility of a dataset for analyses, semantic information can be exploited for attacks such as \gls{tul} \cite{Tu2019,marc2020}.
Semantic information can also be added to a dataset retrospectively, e.g., by matching the location points of a trajectory against semantic datasets, such as OpenStreetMap \cite{OpenStreetMap}.

Protection mechanisms for the release of trajectory datasets target two different scenarios.
In the first scenario, a dataset of multiple trajectories is released, each trajectory represents one record of the dataset. 
Second, one trajectory represents a database and each location can be considered as one record.
Moreover, protection mechanisms either rely on \emph{$k$-anonymity} and related privacy notions, or on \emph{differential privacy}.
$K$-Anonymity follows the intuition of hiding one user in a crowd of users. 
While this class of privacy notion is very intuitive, it cannot provide strong privacy guarantees \cite{Chen2020, Jiang2022, Chen2011, Ma2021, Gomes2018}.
Background knowledge can be exploited to derive knowledge from the protected dataset even when a dataset provides $k$-anonymity or a similar notion.
Therefore, much research has focused on differential privacy (cf.\ Appendix~\ref{sec:dp} for details).
The differential private \gls{sdd} \cite{Jiang2013} mechanism is described in the following section, as we use this mechanism for our evaluation.

\subsection{Sampling Distance and Direction}\label{sec:sdd}

The \glsfirst{sdd} mechanism is one approach to publish location trajectories while providing differential privacy.
We consider the \gls{sdd} mechanism as one target for our reconstruction attack because the mechanism can provide significantly better utility than the standard Laplace mechanism \cite{Jiang2013}. \gls{sdd} is considered as baseline in recent literature on protection mechanisms \cite{Chen2020, Niu2020} and follows an intuitive approach.
As motivation for the approach, the authors refer to the publication of trajectories from ships in the Singapore Straits because unprotected trajectories could be utilised by pirates to plan and launch attacks \cite{Jiang2013}.

The authors make the assumption that start and end locations are not vulnerable and can be published without protection.
Starting from the first location, the exponential mechanism \cite{McSherry2008} is used to sample the \emph{distance} and \emph{direction} to the next location.
The parameters of the exponential mechanism are chosen such that the sampling achieves $\varepsilon$-differential privacy.
Moreover, the point defined by the sampled values must lie within a certain distance to the end point, to preserve utility.
Otherwise, the sampling is repeated.
The evaluation shows that the \gls{sdd} mechanism achieves significantly better utility than the Laplace mechanism, especially for smaller values of $\varepsilon$ (and hence, higher privacy levels) \cite{Jiang2013}. 
Hence, we target this particular mechanism for the evaluation of our attack.

\subsection{Problem Statement}\label{sec:problem-statement}
The protection of ship trajectories with the goal to prevent pirate attacks serves as motivation for the \gls{sdd} mechanism \cite{Jiang2013}.
Other authors \cite{Gomes2018} use stalking and interception of individuals as motivation.
For both threats, the attackers do not necessarily require the exact coordinates.
Instead, the pirates might successfully launch an attack as long as they manage to get into line of sight of the target ship, and for a stalker, an approximate route of the victim suffices.

However, through observation of the output trajectories produced by protection mechanisms, we observe structural differences to the genuine input trajectories.
Structural differences include, for instance, that protected trajectories exhibit zigzag patterns not found in genuine ship trajectories or pass very close or over land, which is impossible for real ships (compare Figure~3 and 4 in \cite{Jiang2013}). 

The goal of this article is to highlight the risk posed by a \emph{\glsfirst{RAoPT}} which exploits such characteristics.
We define the reconstruction attack as a method to reduce the distance between the original and the reconstructed trajectories (called \emph{\gls{OR-Distance}}) significantly compared to the distance between the protected and original trajectories (\emph{\gls{OP-Distance}}).
I.e., the locations of the reconstructed trajectory are physically closer to the locations of the original trajectory than the locations of the protected trajectories.
Reducing the \gls{OR-Distance} significantly, potentially to the level that the reconstructed and original trajectory overlap, represents a serious security threat to the users in the released dataset.
For instance, in case of ships, this would allow the planning of pirate attacks \cite{Jiang2013}, or in case of individuals, this information could be used by stalkers \cite{Gomes2018}.
While a perfect reconstruction of a protected trajectory is not realistic, a partial reconstruction yielding a close physical distance or even intersection, represents a serious privacy breach and security threat.

\subsection{Threat Model}\label{sec:threat-model}
For the evaluation of \gls{RAoPT}, we assume different levels of background information known to the adversary executing the attack.

\subheading{Adversary 1: Full Knowledge.}
The strongest adversary with \emph{full knowledge} knows which mechanism with which parameters have been used for the protection of the released trajectories. 
Moreover, this adversary has access to unprotected trajectory data with the same distribution as the target dataset for the training of the attack model. 
This adversary model is the best case for an attack as the model can be trained on data that has the same properties as the real data.
However, the assumptions are very strong and not realistic in the real-world.
We use this adversary model to examine the influence of different parameters on the reconstruction success.

\subheading{Adversary 2: Partial Knowledge.}
The adversary with \emph{partial knowledge} has no access to unprotected data with the same distribution but is aware of the used protection mechanism and parameters.
Hence, the adversary must train the attack model with data from a different trajectory dataset, e.g., with a publicly available dataset.
Using another dataset during training might lower the attack performance because
the trajectories of the training set might not possess the same unique features as the target dataset.
The adversary with partial knowledge is more realistic than adversary \num{1}, as an adversary is unlikely to get access to unprotected data from the same source as the target dataset. 
The assumption that the protection mechanism is public knowledge is not unrealistic as security through obscurity \cite{kerkhoff-principle} is generally discouraged such that the protection method might be published alongside the dataset.
Hence, this adversary targets a real-world scenario.

\subheading{Adversary 3: No Knowledge.}
In the worst-case, the adversary has \emph{no background knowledge}, i.e., they have neither access to a dataset with the same distribution nor any information about the used protection mechanism or its parameters.
Hence, the attack model has to be trained on data that might display different properties than the dataset that shall be attacked.
Accordingly, the reconstruction success will be lower than for the previous two adversaries.
We provide an overview of related work in the following section.

\section{Related Work}\label{sec:related-work}

In this section, we first describe existing attacks on trajectory protection mechanisms, and then, summarise works applying deep learning to the trajectory domain.

\subheading{Existing Attacks.} Shao et al.\ \cite{Shao2020} proposed an attack framework called \emph{iTracker} that recovers original trajectories from a differential private release by exploiting the correlation between multiple trajectories. 
Previous attacks only used the information of a single trajectory and mostly relied on Markov models \cite{Shao2020}.
The iTracker framework is based on a location sparsity matrix and uses two approximation algorithms to converge towards the most probable original trajectories. 
The evaluation of the approach shows that iTracker is able to recover trajectories that are more similar to the original ones than any of the trajectories recovered by related work.
To the best of our knowledge, this framework represents the most effective attack on differential private trajectory publication mechanisms and highlights that differential private mechanisms can be attacked.
While this framework represents the closest work to our approach,
iTracker is only evaluated on Laplace noise-based mechanisms.
As described in Section~\ref{sec:sdd}, the Laplace noise-based mechanisms add substantially more distortion to a trajectory than more advanced approaches such as the \gls{sdd} mechanism.
Hence, recovery of Laplace noise-protected trajectories is less challenging than trajectories from more advanced protection schemes, as confirmed by our evaluation in Section~\ref{sec:eval}.
Unfortunately, a direct comparison of our approach to iTracker was not possible, as we could not get sufficient implementation details from the authors to reproduce their results.
Moreover, iTracker only utilises time and location information and cannot easily be extended to utilising semantic information.
However, semantic properties can be exploited to improve the performance of attack mechanisms \cite{marc2020, Monreale2011, Shi2021, Tu2019}.

\subheading{Deep Learning-based Protection Mechanisms.}
Few approaches exist which apply deep learning to the trajectory privacy domain \cite{Liu2018a,Qu2020,Rao2020,Chen2020}.
However, all these approaches deploy deep learning for the publication of privacy-protected location trajectories.
To the best of our knowledge, no attack mechanism based on deep learning exists.
In 2018, Liu et al.~\cite{Liu2018a} published a vision paper on the usage of \glspl{gan}~\cite{Luo2019} for the privacy-preserving publication of trajectories. 
The goal of the approach, called \emph{trajGAN}, is the generation of synthetic trajectories that are similar enough to authentic trajectories to provide high utility for analyses.
The proposed framework consists of a generator and a discriminator. 
While the generator tries to generate realistic trajectories, the discriminator has the goal to distinguish between synthetic and authentic trajectories. 
These two components learn from each other such that the synthetic trajectories become harder to distinguish from real trajectories with sufficient training.
Rao et al.~\cite{Rao2020} built upon this vision paper and proposed the \emph{LSTM-trajGAN} model, which utilises an \gls{lstm}~\cite{Hochreiter1997} as the main building block for the \gls{gan}.
This deep learning-based publication approach appears to achieve a better utility-privacy trade-off than traditional publication mechanisms.
The evaluation shows that the synthetic trajectories maintain high utility for analyses while achieving a low trajectory user linking accuracy, which is one indicator of the privacy level.
However, the usage of a deep learning model incurs higher computational costs than traditional publications schemes and also requires an initial time-consuming training process.
Additionally, synthetic trajectories are not suitable for all use cases and the approach requires all input data to originate from a rather small geographical area.
Qu et al.~\cite{Qu2020} follow a similar idea and propose a \gls{gan} to create a differential private synthetic dataset to publish location data collected through 5G networks.
Chen et al.~\cite{Chen2020} utilise a \gls{rnn} to predict a noisy dataset from the original dataset.
This noisy dataset is then further processed to release a differential private dataset that hides the original trajectories. 
The methods used by these approaches inspire the design of our model which we introduce in the following section.

\section{Reconstruction Model}\label{sec:attack}

In this section, we introduce the \gls{lstm}-based \gls{RAoPT} model to address our research question \ref{rq1} defined in Section~\ref{sec:intro}: 
\begin{questions}
    \item Can an adversary (partly) reconstruct trajectories from a differential private trajectory release?
\end{questions}
First, we provide an overview of the attack in Section~\ref{sec:overview}.
Second, we give an overview of pre-processing and encoding in Section~\ref{sec:encoding}.
The structure of the \gls{RAoPT} model is described in Section~\ref{sec:model}.
Finally, we provide details on the training process in Section~\ref{sec:training}.

\begin{figure}
    \centering
    \includegraphics[width=\linewidth]{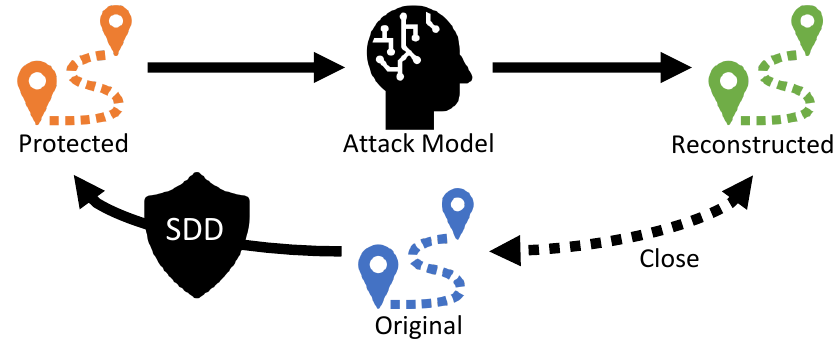}
    \caption{
    Overview of the attack.
    A differential private publication mechanism, e.g.,  the \gls{sdd} mechanism \cite{Jiang2013}, protects the original trajectories.
    These protected trajectories serve as input for the model with tries to reconstruct trajectories that shall be as close as possible to the original versions.
    }
	\Description[Overview of the attack]{
	A differential private publication mechanism, e.g.,  the \gls{sdd} mechanism \cite{Jiang2013}, protects the original trajectories.
	These protected trajectories serve as input for the model with tries to reconstruct trajectories that shall be as close as possible to the original versions.	
    }
	\vspace{-1.5em}
    \label{fig:attack-overview}
\end{figure}

\subsection{Overview}\label{sec:overview}
Differential private publication mechanisms commonly do not take geographical constraints into consideration \cite{Naghizade2020}.
The constraints of real-world trajectories lead to structural differences between the original and the protected trajectories, such as the following: 
Ship trajectories generated by protection mechanisms might pass over land or through shallow waters, while genuine ship trajectories do not exhibit such behaviour.
Moreover, the introduced randomness can lead to atypical zigzag patterns, whereas ships on the open ocean would most likely choose a straight path.
Likewise, realistic taxi or car trajectories have to follow streets, and pedestrians cannot walk through houses and physical barriers.
To address \ref{rq1}, \gls{RAoPT} exploits these characteristics for a partial reconstruction of the original trajectories from the protected release.

A visual overview of the attack is provided in Figure~\ref{fig:attack-overview}.
First, a protection mechanism providing differential privacy, e.g., the \gls{sdd} mechanism (cf.\ Section~\ref{sec:sdd}), protects the \emph{original} trajectories containing private information.
Second, the resulting \emph{protected} trajectories are fed into the \gls{RAoPT} model.
This model returns a \emph{reconstructed} trajectory for each protected trajectory.
The goal of the model is to generate reconstructed trajectories that are close to the original trajectories.
Our attack is successful if the distance between the reconstructed and the original trajectories, called \emph{\gls{OR-Distance}}, is significantly smaller than the distance between the protected and the original trajectories (\emph{\gls{OP-Distance}}).
Before describing the details of the \gls{RAoPT} model in Section~\ref{sec:model}, we describe the trajectory encoding used as input for the model in the following section.

\subsection{Trajectory Encoding and Pre-Processing} \label{sec:encoding}

The general representation of location trajectories is described in Section~\ref{sec:trajectory}.
In this section, we explain how a trajectory is pre-processed and encoded before it can be used as input for the \gls{RAoPT} model.
Trajectories consist of a sequence of locations which can each be composed of multiple properties.
For our attack model, we only utilise the location and time information as these values are the information amount contained in many available datasets \cite{t-drive, Geolife1, foursquareNYC}. 
Through usage of the time information, in particular hour-of-day and day-of-week features, we showcase the capability of the \gls{RAoPT} model to utilise additional semantic knowledge without overstating the attack success by using semantic information that might not be available to a real-world attacker.
Due to the usage of an embedding and feature fusion layer (cf.\ Section~\ref{sec:model}), the model can be extended by further properties, such as \glspl{poi}.
However, the less correlated information per location a dataset provides, the harder a reconstruction becomes.
Therefore, only using time and location represents the worst-case for a reconstruction attack.

\begin{table}
    \begin{center}
        \begin{tabularx}{\columnwidth}{c|c|c|cccc|ccc}
            \toprule
            \textbf{Point}     & \textbf{Latitude}  & \textbf{Longitude} & \multicolumn{4}{c|}{\textbf{Hour of Day}} & \multicolumn{3}{c}{\textbf{DoW}} \\
            \midrule
            \textbf{Dim.} & 1 (float) & 1 (float) & \multicolumn{4}{c|}{24 (binary)}  & \multicolumn{3}{c}{7 (binary)} \\
            1         & -0.80     & 2.34    & 0    & 1    & $\dots$    & 0    & 1       & $\dots$      & 0      \\
            2         & -0.80     & 2.33    & 0    & 1    & $\dots$    & 0    & 1       & $\dots$      & 0      \\
            $\dots$   & $\dots$   & $\dots$   & 0    & 1    & $\dots$    & 0    & 1       & $\dots$      & 0      \\
            38        & 1.23     & 1.45    & 0    & 1    & $\dots$    & 0    & 1       & $\dots$      & 0      \\
            39        & 1.24     & 1.45    & 0    & 1    & $\dots$    & 0    & 1       & $\dots$      & 0     \\
            \bottomrule
        \end{tabularx}
    \end{center}
    \caption{
    Trajectory Encoding.
    The table shows an encoded trajectory consisting of 39 locations.
    Each location contains a latitude, longitude, hour of day and day of week (DoW).
    }
    \label{tab:encoding}
    \vspace{-1.5em}
\end{table}

A matrix represents a trajectory $T$ and each row of this matrix corresponds to one measurement point $t_i$ of the trajectory.
An example of an encoded trajectory is shown in Table~\ref{tab:encoding}.
The location information is represented by latitude and longitude values.
However, instead of using these values directly, we compute the offsets from a central reference point.
I.e., for the reference point $(lat_0, lon_0) = (40.0, 115.0)$, the location point $(39.2, 117.34)$ is represented as $(-0.80, 2.34)$.
This standardisation, motivated by the \gls{lstm}-TrajGAN encoding \cite{Rao2020}, allows the model to better learn the spatial deviation patterns \cite{Rao2020}.
The time information is dissembled into two one-hot encodings.
The hour-of-day is represented by a \num{24}-dimensional binary vector, and the day-of-week by a \num{7}-dimensional vector. 
Both vectors contain exactly one \num{1}-value.
For instance, the first point in Table~\ref{tab:encoding} has been recorded at \SI{2}{am} on a Monday.
Other categorical values could be added to the encoding via similar one-hot encodings. 
For instance, the encoding could be extended by location types such as gym, shopping centre, or medical centre.

Before feeding the encoding into the \gls{RAoPT} model, the latitude, and longitude deviations are scaled through max normalisation \cite{Singh2020}, i.e., all values are divided by the maximal value in the dataset.
Moreover, the trajectory is zero padded to the maximal length expected for any trajectory,
i.e., rows with all values set to \num{0} are appended to the bottom of the matrix.
These padded rows are marked through a masking layer in the model such that they do not have any influence on the training or reconstruction.
After completion of the described pre-processing, the trajectories can be used by the model described in the following section.

\subsection{RAoPT Model}\label{sec:model}

\begin{figure}
	\centering
	\includegraphics[width=\linewidth]{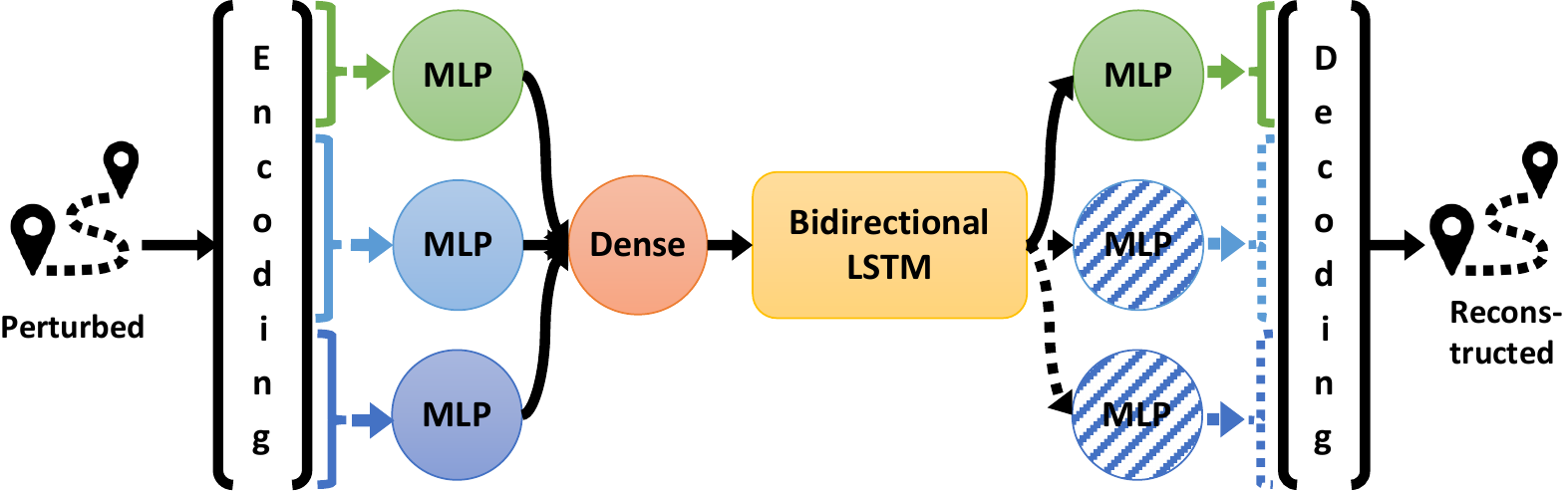}
	\caption{
		Perturbed trajectories are encoded to a vector containing location (green), and time and/or semantic information (light and dark blue).
		These features are embedded by MLPs, concatenated in a feature fusion layer, and fed into the bidirectional \gls{lstm} layer, followed by the output layers.
		Finally, the output is decoded to the reconstructed trajectories.
		The dashed MLPs are optional and not evaluated.
	}
	\Description[Model Overview]{
		Perturbed trajectories are encoded into a vector containing location information, and time and/or semantic information.
		These features are embedded by MLPs, concatenated in a feature fusion layer, and fed into the bidirectional \gls{lstm} layer.
		Finally, the output of the \gls{lstm} is decoded into a vector from which the reconstructed trajectory can be recovered.
		The dashed output MLPs are optional and not used in our evaluation.
	}
	\vspace{-1.5em}
	\label{fig:model}
\end{figure}

The generated encodings can be used for training of the model, or for reconstruction.
Figure~\ref{fig:model} shows the structure of the \gls{RAoPT} model.
Initially, the protected input trajectories are encoded and pre-processed as described in the previous section.

\subheading{Masking Layer.}
The input is masked through a \emph{masking layer} in order to avoid the influence of the padded points on the output.

\subheading{Embedding Layer.}
Next, the input is split into three separate features: \emph{location information} (green), i.e., longitude and latitude, \emph{hour-of-day} (light blue), and \emph{day-of-week} (dark blue).
In case the trajectories include semantic information, more features could be added here.
We treat the hour-of-day and day-of-week as two examples for our model's capability to add semantic information.
Any other features can be added in the same way by passing different parameters during the model's initialisation. 
We decided not to include further semantic features in our evaluation because additional information potentially improves the model's training.
Thus, the evaluated setting represents the worst-case and allows for better generalizability as not every dataset contains additional semantic information.
The model embeds each feature separately through a \gls{mlp} consisting of a dense layer followed by a \gls{relu} activation function.
The \glspl{mlp} use the same weights for all points of the trajectory (realised through a \emph{TimeDistributed layer}).
The units of the dense layer depend on the embedded feature.
For the location information, we use \num{64} units, as this value has worked well for the \gls{lstm}-TrajGAN approach \cite{Rao2020}.
For hour and day, we use the same number of units as the size of the encoding, i.e., \num{24} and \num{6} units.

\subheading{Feature Fusion.}
Then, a dense layer with \num{100} units and a \gls{relu} activation function fuses the concatenation of the embeddings.

\subheading{\gls{lstm} Layer.}
Consecutively, we feed the output of the feature fusion into a bidirectional \gls{lstm} layer with \num{100} units. 
This layer produces one output for each point of the trajectory.

\subheading{Output Layer.}
The output of the bidirectional \gls{lstm} is processed through separate \glspl{mlp} which are again applied to each slice of the sequence with the same weights. 
Each of the \glspl{mlp} consists of a dense layer followed by a \gls{tanh} activation for numerical outputs, such as latitude/longitude, or by a softmax activation in case of categorical output features.
As the considered protection mechanisms only perturb the location information, we use two \glspl{mlp} with one unit each to generate the reconstructed latitude and longitude.
If the model should also reconstruct other information such as timestamps, further \glspl{mlp} can be added. 
However, adding information which was not perturbed and does not require reconstruction is not beneficial as it distracts the model's from the important values.
Moreover, we scale the outputs for latitude and longitude with the inverse scale factor used during pre-processing (cf.\ Section~\ref{sec:encoding}) as the outputs of \gls{tanh} ranges from \num{-1} to \num{1}.

\subheading{Post-Processing.}
The outputs of the actual model have a similar format to the encoding after the pre-processing (cf.\ Section~\ref{sec:encoding}). 
To retrieve useful reconstructed trajectories, the pre-processing steps have to be inverted.
First, the reference point is added to convert the spatial deviations into absolute latitude and longitude values.
Second, the padded points are removed such that the reconstructed trajectory has the same length as the protected trajectory.
Finally, the resulting encoding can be decoded into a trajectory. 
Thereby, any unperturbed information that was not reconstructed can be added back into the trajectory, e.g., the timestamp if only the locations were reconstructed.

\subheading{Loss Function.}
We implemented a custom loss function which computes the \gls{mae} of the Euclidean distance between the output and the ground truth trajectories used during training. 
However, instead of computing the Euclidean distance directly by treating latitude and longitude as coordinates, we compute the haversine distance \cite{haversine-formula} between each pair of locations of the compared trajectories.
For trajectories that have been perturbed with very large amounts of noise (which can be identified by invalid latitude or longitude values), we use the standard \gls{mse} loss function instead because our custom loss function requires valid latitudes and longitudes.
The loss function concludes the description of our \gls{RAoPT} model. 
We provide implementation details and information on hyperparameters in Section~\ref{sec:eval:implementation}.

\subsection{Training}\label{sec:training}

Before the model can be used for the reconstruction of trajectories, it needs to be trained. 
For the generation of training data, the adversary uses trajectories they have access to, for instance, openly available datasets such as T-Drive \cite{t-drive}, GeoLife \cite{Geolife1}, or Foursquare \cite{foursquareNYC}.
The influence of using trajectories with different distribution than the target dataset is evaluated in Section~\ref{sec:eval:results-2}.
The adversary perturbs these trajectories with an available protection mechanism. 
In the best case, they know about the used protection mechanism of the target dataset to attack and use the same mechanism (cf.\ the threat model in Section~\ref{sec:threat-model}).
These generated pairs of original and protected trajectories serve as training data. 
The protected trajectories represent the model's input, while the original trajectories serve as ground truth.
Then, the trained model can be used to reconstruct trajectories from the target dataset.
We evaluate the effectiveness of our attack in the following section.

\section{Evaluation}\label{sec:eval}

In this section, we present the results of our \gls{RAoPT} model's evaluation.
We begin with a description of the used T-Drive and GeoLife datasets in Section~\ref{sec:datasets}, followed by the applied pre-processing in Section~\ref{sec:eval:dataset-processing}.
In Section~\ref{sec:eval:protection}, we provide information about the protection mechanisms which we consider for the evaluation.
Section~\ref{sec:eval:metrics} contains information about the used metrics and Section~\ref{sec:eval:implementation} provides implementation details of the \gls{RAoPT} model.
Finally, we present and discuss the results of our measurements in Section~\ref{sec:eval:results}.

\subsection{Datasets Description}\label{sec:datasets}

To verify the generalizability of the attack, we base our measurements on two datasets. 
First, the T-Drive \cite{t-drive} dataset, which consists of the trajectories of \num{10357} taxis collected over one week in the area of Beijing. 
Second, the GeoLife \cite{Geolife1} dataset, which contains the trajectories of \num{182} users with different modes of transportation collected throughout a period of three years.
While the T-Drive dataset only contains trajectories of similar types, i.e., cars on a street, the GeoLife dataset is more diverse as it contains walking, hiking, running, cycling, driving, and even flights.
The usage of datasets with such different properties allows us to examine the behaviour of our attack in different settings.
Both datasets contain latitude, longitude, and timestamp information. 
Moreover, both datasets allow attributing each trajectory to a certain user (or taxi).
In addition, the GeoLife dataset contains the altitude information for each point. 
However, due to consistency, we do not utilise these values. 
Neither do we enrich the datasets with any semantic knowledge.
Before using the trajectories for our evaluations, we perform a pre-processing step which we describe in the following section.

\begin{figure*}
    \centering
    \includegraphics[width=\linewidth]{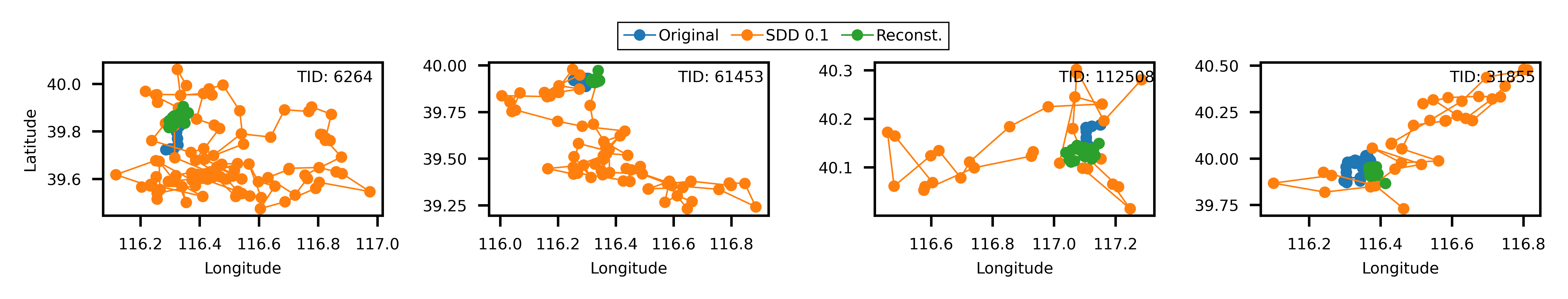}
    \caption{
    Example Trajectory Reconstruction. 
    The figures show four randomly chosen trajectories from the T-Drive dataset (original), after protection with the SDD mechanism ($\varepsilon = 0.1$), and after reconstruction by the \gls{RAoPT} model.
    }\label{fig:examples}
\Description[Example Trajectory Reconstruction]{
	The figures show four randomly chosen trajectories from the T-Drive dataset (original), after protection with the SDD mechanism ($\varepsilon = 0.1$), and after reconstruction by the \gls{RAoPT} model.
	The reconstructed trajectories look significantly more similar to the original trajectories than the protected ones. 
	In all cases the reconstructed trajectories show a large overlap with the original trajectories while showing a more similar density compared to the protected trajectories.

}
	\vspace{-1.5em}
\end{figure*}

\subsection{Pre-Processing}\label{sec:eval:dataset-processing}

We undertake a pre-processing step to sanitise the datasets, as clean data is crucial for good deep learning results \cite{cleanML}.
First, we remove outliers by deleting all location points which lie outside a bounding box defined by the \SI{99}{\%} percentile for T-Drive and by the \SI{95}{\%} percentile for GeoLife.
We use a lower percentile for the GeoLife dataset because it contains locations on other continents which cannot be handled with the reference point approach described in Section~\ref{sec:encoding}.
Second, duplicates are dropped, i.e., locations with the same timestamp.
In case both duplicates refer to the same location, the second point is removed.
In case the two duplicates correspond to the same timestamp, but the locations differ, we assume that the point with a larger distance to the previous and following location is the outlier which can be removed.
Third, speed outliers are removed.
As described in Section~\ref{sec:eval:protection}, we utilise the \gls{sdd} mechanism for our evaluation.
However, the mechanism requires an upper bound on the speed of any user in the trajectory dataset to be defined. 
Therefore, we drop all locations that require that a user has travelled at a speed that is faster than the \SI{99}{\%} percentile for the dataset.
For the T-Drive dataset, all points indicating a speed over \SI{90}{\km\per\hour} are dropped, and for GeoLife all speeds over \SI{100}{\km\per\hour}.

Both datasets contain trajectories of varying lengths. 
The T-Drive dataset only contains one trajectory per taxi, which spans the time period of a week, while the GeoLife dataset contains multiple trajectories of different lengths per user.
We define a trajectory as the locations of one uninterrupted trip, for instance, one workday of a taxi-driver in case of the T-Drive dataset.
To depict this, we split the trajectories based on a time gap of \SI{11}{\min} (To include trajectories with one GPS reading every \SI{10}{\min}) for T-Drive, and \SI{20}{\s} (the \SI{99}{\%} percentile) for GeoLife.
Finally, we remove trajectories that are shorter than \num{10} locations, as they do not contain much information content. We also remove trajectories longer than a threshold of \num{100} points for T-Drive and \num{200} points for GeoLife.
The reason for defining an upper threshold is that the deep learning model requires padding of all trajectories to the same size for efficient training and reconstruction.
After this pre-processing, the processed T-Drive dataset contains \num{163006} trajectories, and the processed GeoLife dataset \num{90146}, respectively.
Now, the trajectories can be protected by a differential private publication mechanism, as described in the following section, to generate the inputs for the \gls{RAoPT} model.

\subsection{Protection Mechanisms}\label{sec:eval:protection}

Our attack targets differential private trajectory publication mechanisms.
A number of approaches \cite{Niu2014,Li2017,Liu2021a,Hua2015,Cao2021,Jiang2013} have been proposed, but for most, no implementation is openly available such that a time-consuming re-implementation is required.
Related works based their evaluation on the simple Laplace mechanism (cf.\ Section~\ref{sec:related-work}), however, we assume that reconstruction from more sophisticated approaches adding less total noise is a harder problem.
To achieve realistic results, we decided to consider two different protection methods.
First, we use a simple Laplace noise based mechanism as a baseline. 
In particular, we consider the \textbf{CNoise mechanism} defined by Jiang et al.\ \cite{Jiang2013} because it is the best performing Laplace noised-based mechanism examined in the paper. 
Second, we utilise the \textbf{\gls{sdd} mechanism} \cite{Jiang2013} (cf.\ Section~\ref{sec:sdd}) which is considered a state-of-the-art protection mechanism \cite{Chen2020, Niu2020}. 

We implemented both mechanisms as close as possible to the descriptions in the original paper, however, had to make a few minor changes to the \gls{sdd} mechanism.
For long trajectories, the mechanism frequently got stuck on line \num{11} of the original algorithm definition (Algorithm 5 in \cite{Jiang2013}).
To avoid infinite runtime, we restart the entire algorithm in case the inner loop does not terminate after \num{1000} runs.
Moreover, after completion of the standard mechanism, we also perturb the start and end point by sampling a distance and direction from the second (last) point, as not in every scenario start and end point are public knowledge.
As we restart the entire algorithm in case of the first modification and add further perturbation in case of the second, both modifications do not lower the level of differential privacy provided by the mechanism.

Both mechanisms require a sensitivity $M$ for each dataset to add the appropriate amount of noise to achieve differential privacy.
We choose $M$ to be \SI{16500}{\m} as this is the sensitivity computed for the T-Drive dataset by multiplying the maximal speed (\SI{90}{\km\per\hour}) with the sampling rate (\SI{11}{\min}).
The GeoLife dataset would allow for a lower sensitivity due to the finer sampling rate.
Due to consistency we choose the larger value \SI{16500}{\m} for all measurements, as a larger choice is valid while a too low sensitivity breaks differential privacy guarantees.
Furthermore, the locations provided in latitude $lat$ and longitude $lon$ need to be converted into Cartesian coordinates before the application of the protection mechanisms.
For simplicity, we use offset coordinates from a central point $(lat_0, lon_0)$, as both pre-processed datasets only contain locations within a certain bounding box.
With  \SI{111319.44}{\metre} as the average distance between two degrees of latitude, we use the following formula:
\begin{equation*}
    \begin{aligned}
        x&=&\num{111319.44} * \cos{lat_0} * (lon - lon_0)\\
        y&=&\num{111319.44 } * (lat - lat_0)
    \end{aligned}
\end{equation*}
After application of the mechanism, we transform the locations back into latitude and longitude values.
In our measurements, we consider different values for the privacy parameter $\varepsilon$ (cf.\ Section~\ref{sec:dp}) which, in practice, usually takes values between \num{0.01} and \num{10} \cite{Erlingsson2014}.
For the measurements with the \gls{sdd} mechanism, we focus on values from this range.
For the CNoise mechanism, we additionally consider $\varepsilon = \num{100}$ as the mechanism is faster to apply, and the generated trajectories are very close to the original ones, which is the worst-case scenario for our attack.

\subsection{Metrics}\label{sec:eval:metrics}

The goal of the reconstruction attack is to minimise the physical distance between the locations of the original and the reconstructed trajectories.
To measure this distance, we consider three metrics: (1) the Euclidean distance, (2) the Hausdorff distance, and (3) the Jaccard index of the trajectories' convex hulls.
Both the Euclidean distance \cite{Jiang2013, Shao2020} and the Hausdorff distance \cite{Rao2020,Shao2020,Ma2021,Hua2015,Li2017} have been widely used to measure the distance of trajectories.
To compute the distance between two locations defined through latitude and longitude, we use the haversine formula \cite{haversine-formula}.
To simplify the discussion, we focus on metrics (1) and (2) in the following.
We explain the Jaccard Index and discuss its suitability for our evaluation in Appendix~\ref{sec:a:jaccard}, and record all results in Table~\ref{tab:all-results} in the appendix.

\subsection{Implementation}\label{sec:eval:implementation}
We implemented the \gls{RAoPT} model presented in Section~\ref{sec:model} with Keras \cite{keras} contained in TensorFlow 2.4.1 \cite{tensorflow} using Python 3.9.
Our implementation relies on NumPy \cite{numpy} in version 1.19.2, and pandas \cite{pandas} in version 1.4.2.
We compute the haversine distance with the haversine library \cite{haversine}.
The model uses the Adam optimiser \cite{adam-optimizer} with a learning rate of \num{0.001}.
We choose a batch size of \num{512}, trained our model for maximally \num{500} epochs, but terminated the training process with an early stop patience of \num{50} epochs.
For other hyperparameters, the default values of the libraries are used.
In test cases using one dataset, we use \num{5}-fold cross-validation, i.e., we perform \num{5} runs, train on \SI{80}{\%} of the dataset and test on the other \SI{20}{\%}. 
For the test cases using different datasets,we perform \num{5} independent runs on the entire datasets.

\begin{figure}
    \centering
    \includegraphics[width=\linewidth]{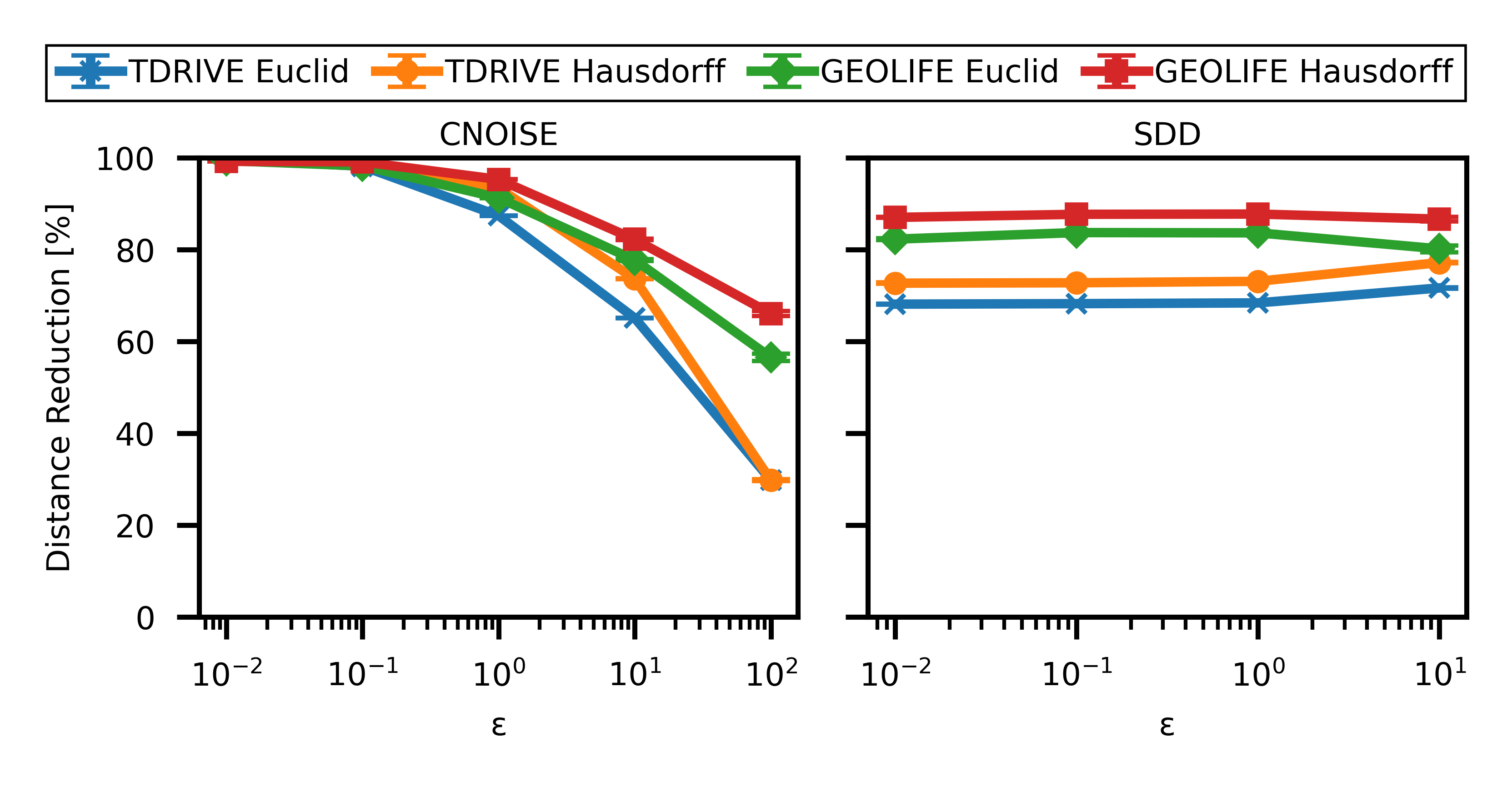}
    \caption{
    The plot shows the percentage reduction of the \gls{OR-Distance} compared to the \gls{OP-Distance}.
    The left plot shows the results for CNoise, the right plot for the SDD mechanism.
    }
    \label{fig:case1}
    \Description[Distance Reductions]{The plot shows the percentage reduction of the \gls{OR-Distance} compared to the \gls{OP-Distance}.
    	The left plot shows the results for the CNoise protection mechanism, the right plot for the SDD mechanism. 
		While the left plot shows a decreasing curve starting at nearly \SI{100}{\%} for $\varepsilon = 0.01$ and ending at less than \SI{40}{\%} for $\varepsilon = 100$, the curves for the SDD mechanism stay nearly constant at over \SI{80}{\%} for the GeoLife dataset and around \SI{70}{\%} for the T-Drive dataset.
}
    \vspace{-1.5em}
\end{figure}

\subsection{Results}\label{sec:eval:results}

In this section, we provide the results of our measurements.
All bar plots show the average distance reduction (Formula in Appendix~\ref{sec:a:reduction-computation}) and the \SI{99}{\%} confidence intervals as error bars.
We omit the results for the Jaccard index in the plots to make the appearance clearer.
A table containing the results of all performed measurements can be found in Appendix~\ref{sec:a:all-results}.
We consecutively discuss the adversaries defined through our threat model in Section~\ref{sec:threat-model}, beginning with adversary~1 in Sections~\ref{sec:eval:results-tdrive}-\ref{sec:eval:results-mechanism}, followed by adversary~2 in Section~\ref{sec:eval:results-2}, and finally, adversary~3 in Section~\ref{sec:eval:results-3}.
In addition, we provide runtime measurements in Appendix~\ref{sec:eval:runtime}.

\subsubsection{Adversary 1: T-Drive Dataset.}\label{sec:eval:results-tdrive}

First, we examined the performance of the \gls{RAoPT} model on the T-Drive \cite{t-drive} dataset with different protection mechanisms.
Figure~\ref{fig:case1} displays the average reduction of the \gls{OR-Distance} compared to the \gls{OP-Distance} for both protection mechanisms and different values of $\varepsilon$.
In case of the \gls{sdd} mechanism, the reconstruction attack can reduce the distance to the original trajectories by over \SI{68}{\%} %
for all choices of $\varepsilon$.
The Jaccard index is increased through reconstruction by over \SI{180}{\%} on average.
The $\varepsilon$ parameter has only limited influence on the outputs of the mechanism due to the way the mechanism is designed.
This finding is in-line with the results of the original paper \cite{Jiang2013}.

In case of the CNoise mechanism being used, the reconstruction even reduces the distances by far above \SI{80}{\%} for $\varepsilon \leq 1$.
For $\varepsilon = 10$, the distance is still reduced by ~\SI{65}{\%} (Euclidean distance) and ~\SI{74}{\%} (Hausdorff distance), respectively.
For $\varepsilon = 100$, the reconstructed trajectories are only ~\SI{30}{\%} closer to the original trajectories.
In this setting, the CNoise mechanism barely perturbs the trajectories, such that the protected trajectories are already very close to the originals.
However, such a high value for $\varepsilon$ is very unlikely to be used in the real world \cite{Erlingsson2014} as it cannot provide much privacy.
The  average Jaccard index is increased in all cases, from \SI{11}{\%} for $\varepsilon = 100$ to an increase by factor \num{430573} for $\varepsilon = 0.01$.

By means of illustration, four randomly chosen trajectories, their \gls{sdd} ($\varepsilon = 0.01$) protected versions, and the reconstruction results are displayed in Figure~\ref{fig:examples}.
For all these trajectories from the T-Drive dataset, the reconstructed trajectories are not only significantly closer to the original ones, but the structure, e.g., in terms of density and space between trajectories, is much more similar.
This finding is captured by the significant increases of the Jaccard index which resembles the similarity of the trajectories' activity spaces.
An adversary with the intention to intercept the user of the original trajectories has a reasonable chance of success by using the reconstructed trajectories.
In the following section, we perform the reconstruction attack on our second dataset.

\begin{table}
	\begin{center}
    	\begin{tabularx}{\columnwidth}{l|cYYYY}
        	ID & Mechanism & $\varepsilon$ Train & $\varepsilon$ Test & Euclidean & Hausdorff 
        	\\ \midrule 
        	27 & CNoise & 1.0 & 10.0 & \SI{24.3}{\%} & \SI{46.2}{\%} \\
        	28 & CNoise & 10.0 & 1.0 & \SI{72.5}{\%} & \SI{79.3}{\%} \\
        	29 & SDD & 0.1 & 1.0 & \SI{68.4}{\%} & \SI{73.1}{\%} \\
        	30 & SDD & 1.0 & 0.1 & \SI{68.3}{\%} & \SI{72.8}{\%} \\
    	\end{tabularx}
    \end{center}
	\caption{
	Except for the varied $\varepsilon$, the same parameters have been used for train and test set based on the T-Drive dataset.
	The table shows the average distance reductions.
	}\label{tab:epsilon-influence}
	\vspace{-1.5em}
\end{table}

\subsubsection{Adversary 1: GeoLife Dataset.} \label{sec:eval:results-geolife}

To show that \gls{RAoPT} is generally applicable to different datasets, we performed the same measurements on the GeoLife \cite{Geolife1} dataset.
The results are also displayed in Figure~\ref{fig:case1}, alongside the results for the T-Drive dataset.
The figure depicts that the reduction of the distances is very similar to the T-Drive results, just slightly higher in all cases.
We presume that this is caused by the GeoLife dataset being more diverse and hence, the training leads to a more robust model which generalises better to the test set.
The average Jaccard indices can even be increased by over \SI{4900}{\%} in all cases except for CNoise with $\varepsilon = 100$, where the increase is still larger than \SI{600}{\%}.
Next, we examined the influence of training and testing on datasets protected with different $\varepsilon$.

\begin{table}
    \begin{center}
        \begin{tabularx}{\columnwidth}{l|YYYYY}
        ID & Train & Test & $\varepsilon$ & Euclidean & Hausdorff 
        \\ \midrule
        31 & CNoise & SDD & 1.0 & \SI{27.7}{\%} & \SI{44.9}{\%} \\
        32 & SDD & CNoise & 1.0 & \SI{53.0}{\%} & \SI{70.3}{\%} \\
        \end{tabularx}
    \end{center}
    \caption{
    All cases use the T-Drive dataset.
    Except for the protection mechanism, the same parameters have been used for train and test set, including the same $\varepsilon$ value.
    }\label{tab:mechanism-influence}
    \vspace{-1.5em}
\end{table}

\subsubsection{Adversary 1: Influence of Different $\varepsilon$.}\label{sec:eval:results-epsilon}

In the real world, an adversary might not know the exact parameters used for the released dataset.
Therefore, we examined four cases in which the reconstruction model is trained on trajectories that are protected with $\varepsilon$ set to a different value than the test set.
We consider \num{0.1} and \num{1} for the \gls{sdd} mechanism, as these are in the middle of the common range of values (cf.\ Section~\ref{sec:eval:protection}), and \num{1} and \num{10} for CNoise, as lower values for CNoise lead to a level of perturbation that renders the protected trajectories meaningless.
Table~\ref{tab:epsilon-influence} shows the average distance reduction through the reconstruction attack.
Measurement 27 shows that the reconstruction of trajectories protected with CNoise and $\varepsilon = 10$ performs worse when trained on a dataset with $\varepsilon = 1$.
This is caused by the fact that CNoise($\varepsilon=1$) adds substantially more perturbation to the trajectories compared to CNoise $\varepsilon=10$.
Therefore, the model modifies the input trajectories more than necessary.
The opposite test case 28 reduces the distances only \SI{15}{\%} less than training on a dataset protected with the same parameters.

In case of the \gls{sdd} mechanism, the results are very similar to the cases in Section~\ref{sec:eval:results-tdrive} with training on the same parameters.
Both the Euclidean and the Hausdorff distance can still be reduced by over \SI{68}{\%}, and the average Jaccard index can be increased by over \SI{180}{\%}.
This observation can be explained by the fact that the outputs of the \gls{sdd} mechanism are barely affected by the choice of $\varepsilon$ (cf.\ Section~\ref{sec:eval:results-tdrive}).
In some cases, not only the parameter, but the entire used protection algorithm might be unknown to the adversary. 
We examine this situation in the following section.

\subsubsection{Adversary 1: Influence of Different Mechanism.}\label{sec:eval:results-mechanism}

Table~\ref{tab:mechanism-influence} shows the distance reduction for two cases where the training dataset is protected with a different mechanism than the test set.
We fix $\varepsilon$ to \num{1} to determine the influence of a varied mechanism only.
While the reconstruction attack still reduces the \gls{OR-Distance} by at least \SI{27}{\%} %
in comparison to the \gls{OP-Distance}, the reconstruction success is significantly smaller than training and testing on datasets protected with the same mechanism.
This might be caused by the different amounts of perturbation added by different mechanisms, and by different characteristics of the algorithms.
While the protection mechanism being public knowledge is not an unrealistic assumption (cf.\ Section~\ref{sec:threat-model}), and adversary has rarely access to training data from the same source as the target dataset.
Therefore, we consider the case of different datasets in the following section.

\subsubsection{Adversary 2: Dataset Transfer.}\label{sec:eval:results-2}

\begin{figure}
    \centering
    \includegraphics[width=\linewidth]{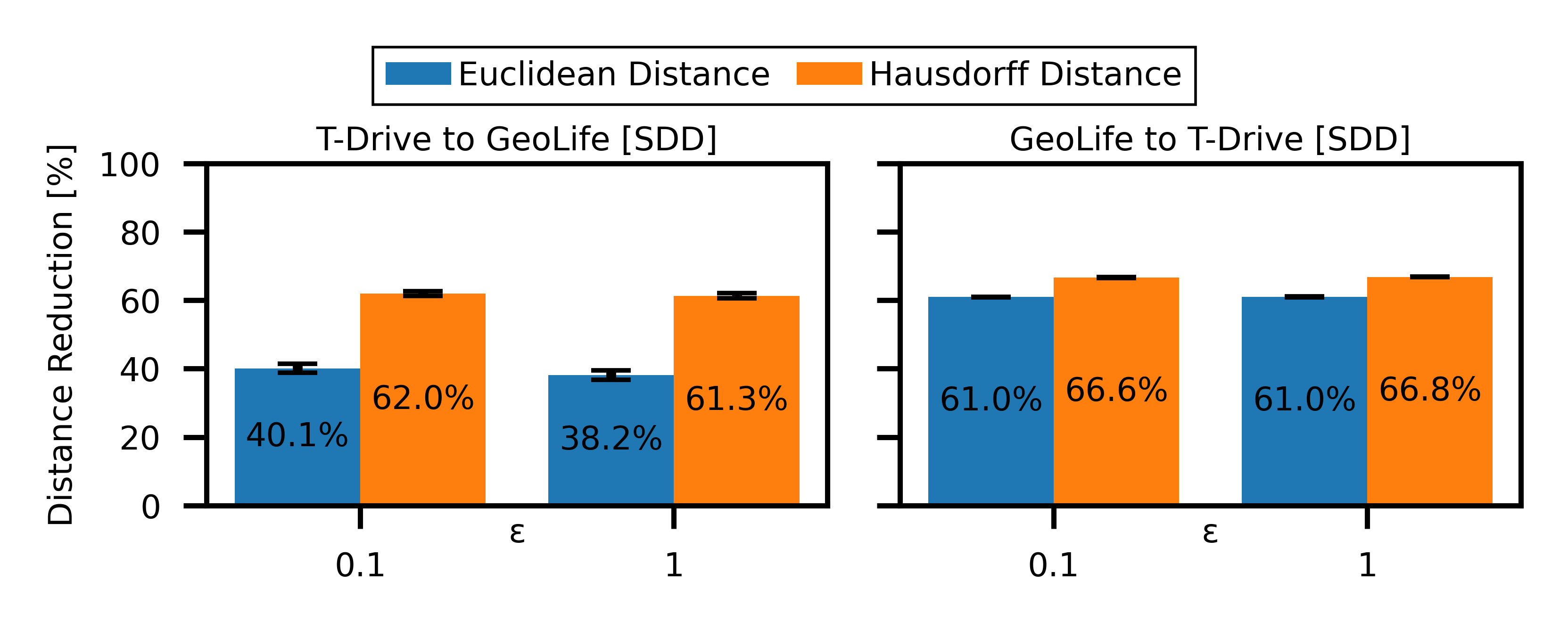}
    \caption{
    Adversary 2:
    The figure shows the results for the transfer from one dataset to another. 
    All test cases use the \gls{sdd} mechanism as protection with different choices for $\varepsilon$.
    }
    \label{fig:transfer_sdd}
    \Description[Distance reductions for adversary 2]{
	The figure shows the distance reduction of the reconstruction attack after training on one dataset and testing on the other. 
	Both test cases use the \gls{sdd} mechanism as protection with different choices for $\varepsilon$.
	The Euclidean Distance shows a slightly smaller reduction than the Hausdorff distance in all cases.
	For the Hausdorff distance, all reductions lie over \SI{60}{\%}, while the Euclidean distances can be reduced by around \SI{40}{\%} for the T-Drive to GeoLife transfer and around \SI{60}{\%} for the opposite direction.
}
    \vspace{-1.5em}
\end{figure}

A real-world adversary might not have access to trajectories of the same distribution as the trajectories they try to attack.
Therefore, the \gls{RAoPT} model needs to be trained on a different dataset, e.g., a publicly available dataset. 
To investigate the attack performance in such a setting, we performed measurements training our model on one dataset and using the other dataset as the test set.
The results for protection with the \gls{sdd} mechanism are shown in Figure~\ref{fig:transfer_sdd}.
The corresponding measurements for CNoise are provided in Appendix~\ref{sec:a:ad2cnosie}.

Figure~\ref{fig:transfer_sdd} still shows reductions of approx. \SI{61}{\%} for the Euclidean and approx. \SI{67}{\%} for the Hausdorff distance, when transferring from GeoLife to T-Drive. 
The opposite direction is less successful with approx. \SI{40}{\%} reduction of the Euclidean and approx. \SI{61}{\%} for the Hausdorff distance.
This finding could be caused by the higher diversity of the GeoLife trajectories which include trajectories that are similar to the T-Drive trajectories, while the T-Drive dataset does not contain all transportation modes of the GeoLife dataset.
The average Jaccard index shows increases in all mentioned cases.

If the CNoise ($\varepsilon = 1$) %
mechanism is used for protection, the reconstructed trajectories show an over \SI{90}{\%} reduced Hausdorff distance.
The attack reduces the Euclidean distance by \SI{76.7}{\%} (T-Drive to GeoLife) %
and \SI{82.9}{\%} (GeoLife to T-Drive), %
respectively.
While for CNoise ($\varepsilon = 10$), %
the transfer from T-Drive to GeoLife cannot reduce the Euclidean distance at all, the other direction %
can even achieve a reduction by \SI{48}{\%} and \SI{59}{\%} for the Euclidean and Hausdorff distances, respectively.

In conclusion, these measurements indicate that \gls{RAoPT} is also successful for adversary 2, who does not have access to optimal training data.
Therefore, this evaluation highlights the real-world danger posed by the proposed reconstruction attack.
To consider an even more realistic scenario, we provide measurements assuming no background knowledge by the adversary in the following section.

\subsubsection{Adversary 3: No Background Knowledge.}\label{sec:eval:results-3}

\begin{figure}
    \centering
    \includegraphics[width=\linewidth]{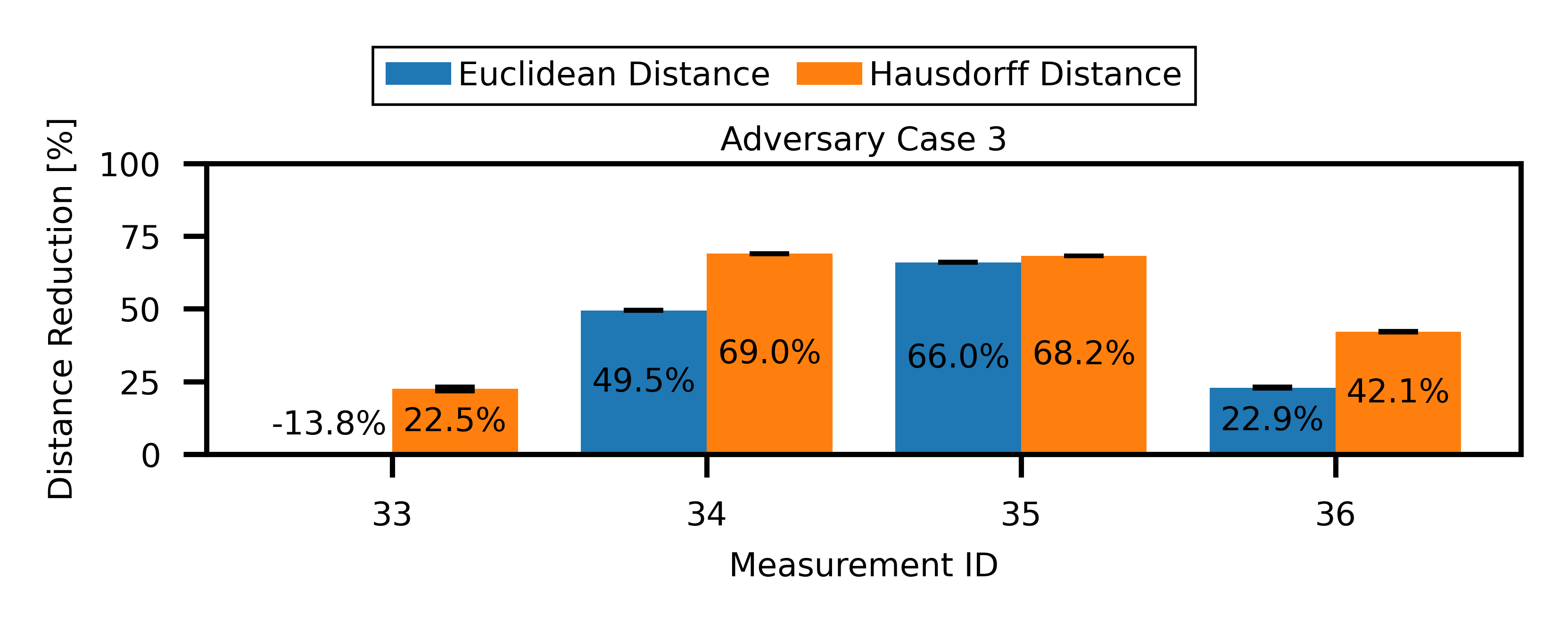}
    \caption{
    Adversary 3:
    The figure shows 4 measurements for which all considered parameters have been modified.
    }
 	\Description[Distance reductions for adversary 3]{
	The figure shows that in case 33 the reconstruction is not successful at all, while both distances can be reduced by \SI{49}{\%} or more for case 34, by over \SI{66}{\%} in case 35 and over \SI{20}{\%} for case 36.
 }
    \label{fig:worst-case}
    \vspace{-1.5em}
\end{figure}

The worst-case scenario for an adversary assumes that no background knowledge about the dataset or protection is known.
In particular, the training dataset does not match the testing dataset in terms of properties, the used protection mechanism is unknown as well as the choice of the $\varepsilon$ parameter.
We picked four such cases, differing both the dataset and the protection mechanism for the training dataset, as well as different common values for $\varepsilon$ of either \num{0.1} or \num{1}.
The concrete specifications are shown in Table~\ref{tab:worst-case}, and the results in Figure~\ref{fig:worst-case}.

In case 33, the reconstruction is not successful, as it combines the transfer from CNoise protected trajectories to \gls{sdd} protected trajectories and the transfer from a larger $\varepsilon$ to a smaller value. 
Case 34, which switches the direction of dataset, mechanism and parameter transfer, on the other hand, allows a distance reduction of \SI{49}{\%}/\SI{59}{\%} (Euclidean/Hausdorff).
Case 35 shows even better reconstruction success with a \SI{66}{\%} reduction of the Euclidean and \SI{68}{\%} for the Hausdorff distance.
Finally, case 36 only allows for a limited reduction of \SI{23}{\%}, and \SI{42}{\%}, respectively.
Notably, the average Jaccard index is significantly increased through reconstruction for cases 34 and 35, stays nearly unaltered for case 33 (\SI{8}{\%} increase) but even shows a decrease for case 36.

These results indicate that even an adversary without background knowledge can execute a successful reconstruction attack and in this way, harm the privacy of contained users.
Moreover, the results show that certain parameter choices for the training set can be helpful for a more successful reconstruction.
First, training on trajectories protected with less perturbation than the target dataset appears to yield better results, i.e., training on a set with a larger $\varepsilon$ parameter or \gls{sdd} instead of CNoise. 
Hence, an adversary without knowledge about the protection mechanism should train on a dataset protected with a mechanism adding limited noise.
Second, transferring from a more diverse training set seems to be advantageous, which matches deep learning best practices.
By choosing appropriate parameters and using a training dataset with characteristics close to the target dataset, an adversary can successfully reconstruct trajectories from the protected set.
This reconstruction represents a threat to all users whose data is contained in the released and seemingly protected dataset.

\begin{table}
\begin{tabularx}{\columnwidth}{l|cccccc}
ID & DS Train & DS test &  Train & Test & $\varepsilon$ Train & $\varepsilon$ Test\\ \midrule
33	&	T-Drive	&	GeoLife	&	CNoise	&	SDD	&	1.0	&	0.1\\
34	&	GeoLife	&	T-Drive	&	SDD	&	CNoise	&	0.1	&	1.0\\
35	&	T-Drive	&	GeoLife	&	SDD	&	CNoise	&	1.0	&	0.1\\
36	&	GeoLife	&	T-Drive	&	CNoise	&	SDD	&	0.1	&	1.0\\ 
\end{tabularx}
\caption{
Adversary 3:
Specifications of our four worst-case measurements.
}\label{tab:worst-case}
\vspace{-2.5em}
\end{table}

\section{Discussion}\label{sec:discussion}

In this section, we further discuss our findings, mention possible countermeasures, and outline opportunities for future work.
The goal of this article was to investigate research question~\ref{rq1}:
\begin{questions}
   \item Can an adversary (partly) reconstruct trajectories from a differential private trajectory release?
\end{questions}
Considering the results of our evaluation, we can answer this question affirmatively.
The measurements described in the previous section highlight that the proposed \glsfirst{RAoPT} successfully reduces the \gls{OR-Distance} compared to the \gls{OP-Distance} in most cases. 
The results show, that the attack is not limited to a knowledgeable adversary, but an adversary that needs to transfer from one dataset to another can still achieve distance reductions of approx. \SI{60}{\%} or more for most considered protection mechanisms on some datasets.

It is important to understand that our attack does not show a vulnerability of the mathematically proven privacy notion of differential privacy (cf.\ Appendix~\ref{sec:dp}).
Rather than targeting the notion itself, the reconstruction attack targets the concrete mechanism of achieving differential privacy.
Our results do \emph{not} indicate that all differential private publication mechanisms have to be susceptible to our \gls{RAoPT}.
The issue is rather that current protection mechanism do not consider the characteristics of genuine trajectories sufficiently.
Accordingly, the design of improved differential private publication mechanisms is recommended for future work.

As mentioned in Section~\ref{sec:related-work}, a direct comparison to the iTracker \cite{Shao2020} attack is not possible due to missing implementation details.
However, iTracker only targets Laplace perturbation, similar to the CNoise mechanism. 
In particular, iTracker's evaluation considers the Laplace mechanism with $0.1 \leq \varepsilon \leq 0.9$.
As shown in Section~\ref{sec:eval}, the reconstruction of trajectories protected with CNoise and $\varepsilon \leq 1$ allows for very high distance reductions.
For adversary 1 with training and testing on sets with the same parameters, \gls{RAoPT} manages to reduce the distances by \SI{87}{\%} or more.
Also, both worst-case measurements targeting CNoise (cf.\ Section~\ref{sec:eval:results-3}) achieve very high reconstruction rates of \SI{49}{\%} - \SI{69}{\%}.

\subheading{Countermeasures.}\label{sec:countermeasures}
The evaluation results clearly show that mechanisms adding less perturbation are less vulnerable to the reconstruction attack.
Accordingly, the design of protection mechanisms adding a minimal amount of noise while sufficiently protecting privacy combines high utility with a protection against a reconstruction attack.
Approaches such as LSTM-TrajGAN \cite{Rao2020} are by design more robust against reconstruction attacks as the generator of the \gls{gan} is trained to generate trajectories which are indistinguishable from authentic trajectories.
However, as described in Section~\ref{sec:related-work}, the approach is not applicable in all scenarios.
Accordingly, extending similar approaches to further use cases can represent an effective countermeasure against reconstruction attacks.
Apart from that, the reconstruction attack exploits the different characteristics of protected and original trajectories. 
Hence, designing protection mechanisms that produce trajectories with realistic characteristics can effectively counteract reconstruction attacks.

\subheading{Future Work.}\label{sec:future-work}
In future work, we intend to highlight the privacy threat of the attack by comparing the success of a \gls{tul} attack before and after the reconstruction through \gls{RAoPT}.
Due to the usage of location offsets (cf.\ Section~\ref{sec:encoding}), the model can only handle trajectories from a limited geographical area. 
Future work could look into the generalisation of the attack to trajectories with arbitrary locations.
Moreover, the influence of adding semantic features to the trajectories on the reconstruction success could be determined, as semantic information can be exploited for more accurate attacks \cite{Primault2015, Rao2020, marc2020}.
Lastly, the focus of further research should lie on the development of novel privacy-preserving trajectory publication mechanisms, which provide both high levels of utility and privacy, and are not susceptible to reconstruction attacks.

\section{Conclusion}\label{sec:conclusion}
While location trajectories offer huge potential for many use cases such as navigation, marketing, or pandemic control, this datatype is very sensitive because it can reveal religious, political or sexual beliefs. 
Therefore, trajectory datasets require appropriate protection before publication.
Due to its strong theoretical guarantees, differential privacy represents the basis for most recent publication mechanisms.
However, the perturbation caused by these publication mechanisms makes it possible to distinguish published trajectories from authentic trajectories. 
Structural differences, e.g., cars not following roads, can be exploited to partially recover the original trajectories from a differential private publication, and hence, impair the privacy of individuals in the dataset.
To highlight these shortcomings, we propose the \glsfirst{RAoPT}.
The \gls{lstm}-based model can significantly reduce the distance of protected trajectories to the original versions.
In addition to simple perturbation-based protection, we target the more practical \gls{sdd} publication mechanism.
To measure the success of our attack, we compute the reduction of the Euclidean and Hausdorff distances, as well as the increase of the Jaccard index of the convex hull.
On the T-Drive dataset both distances can be reduced by over \SI{68}{\%} %
while the Jaccard index can be increased by over \SI{180}{\%}
for trajectories protected with either protection method and $\varepsilon \leq 1$.
An adversary that has to train the \gls{RAoPT} model on a different dataset can still successfully reconstruct trajectories, as the transfer from the GeoLife to the T-Drive dataset allows for an over \SI{60}{\%} distance reduction considering protection with the \gls{sdd} mechanism and $\varepsilon = 0.1$ or $\varepsilon = 1$, and a \SI{30}{\%} increased Jaccard index. 
These results indicate that a reconstruction attack represents a significant privacy threat to existing trajectory publication mechanisms.
Thus, further research on improved privacy-preserving publication mechanisms for trajectory datasets is required.

\begin{acks}
The authors would like to thank UNSW, the Commonwealth of Australia,
and the Cybersecurity Cooperative Research
Centre Limited for their support of this work.
The authors would like to thank all the anonymous reviewers for their valuable feedback.
\end{acks}

\bibliographystyle{ACM-Reference-Format}
\bibliography{library}

\appendix

\printglossaries

\section{Differential Privacy}\label{sec:dp}

Differential Privacy \cite{Dwork2008} represents one of the central privacy notions used to protect personal information.
Compared to other notions, it is based on strong theoretical guarantees and provides protection even against adversaries with background knowledge.
The main intuition of differential privacy is that the input of any single user or row in a dataset does not significantly change the published result.
Accordingly, participation does not harm any user's privacy as the output is approximately the same with or without their data.
The mathematical definition is as follows \cite{Dwork2008}:

\begin{definition}[Differential Privacy]
	A mechanism $\mathcal{K}$ provides $\varepsilon$-differential privacy if for all datasets $D_1$ and $D_2$ differing in at most one element, and all $S \subseteq Range(\mathcal{K})$ holds
	\begin{equation}
		\mathds{P}[\mathcal{K}(D_1)\in S] \leq e^{\varepsilon}\times \mathds{P}[\mathcal{K}(D_2)\in S]
	\end{equation}
\end{definition}

For example, the mechanism $\mathcal{K}$ might be a function that computes a noisy average over a dataset.
Now if the data of another user is added to the dataset $D_1$ yielding dataset $D_2$, the change of the probabilities for the outputs of $\mathcal{K}$ is bounded depending on $\varepsilon$.
The smaller $\varepsilon$ is chosen, the larger is the provided privacy level.
In literature, common values for $\varepsilon$ range from \num{0.01} to \num{10} \cite{Erlingsson2014}.
The most common way to design a differential private mechanism is the addition of noise to the output, by using the Laplace mechanism \cite{Dwork2006}, Gaussian mechanism \cite{Dwork2013}, or exponential mechanism \cite{McSherry2008}.
While adding noise from a Laplace distribution to location data is a straight-forward way to achieve differential privacy \cite{Shao2020, Jiang2013}, the sensitivity of location information requires that a high level of noise be added to the published trajectories, such that they cannot provide much utility \cite{Jiang2013}.
Therefore, multiple differential private mechanisms specifically targeting trajectories have been designed \cite{Chen2011,Li2017,Liu2021a,Hua2015,Cao2021,Jiang2013}.
One example is the \gls{sdd} mechanism which we describe in Section~\ref{sec:sdd}.

\section{Jaccard Index}\label{sec:a:jaccard}
In addition to the Euclidean and Hausdorff distances described in Section~\ref{sec:eval:metrics}, we also measured the Jaccard index of the trajectories convex hulls for each measurement.
The convex hull of a trajectory can be used to represent the activity space of a user \cite{Lee2016}.
Hence, the Jaccard index of two trajectories' convex hull, which is computed by dividing the intersection through the union of two areas, indicates how close the activity spaces are. 
A Jaccard index of \num{1} means that the activity spaces are identical, while \num{0} implies that the activity spaces do not intersect.
The Jaccard index has not only been used before to measure trajectory closeness \cite{Rao2020}, but is particularly suitable to indicate the threat posed by reconstruction. 
A large Jaccard index for a reconstructed trajectory suggests that an attacker, e.g., a stalker, will find the victim within the activity space of the reconstructed trajectory.
The index is better suited than the intersection itself, as it penalises very large activity spaces which include the original trajectory, e.g., a protected trajectory that spans over a large area due to the high noise.
Such a large area containing the original trajectory is not helpful for an adversary, as the adversary would only learn that the victim is somewhere within the large area.
The smaller the convex hull of the reconstructed trajectory is, the less area needs to be considered/searched by the adversary.
Table~\ref{tab:all-results} states the mean Jaccard index before and after reconstruction for all our evaluation cases.

\begin{figure}
    \centering
    \includegraphics[width=\linewidth]{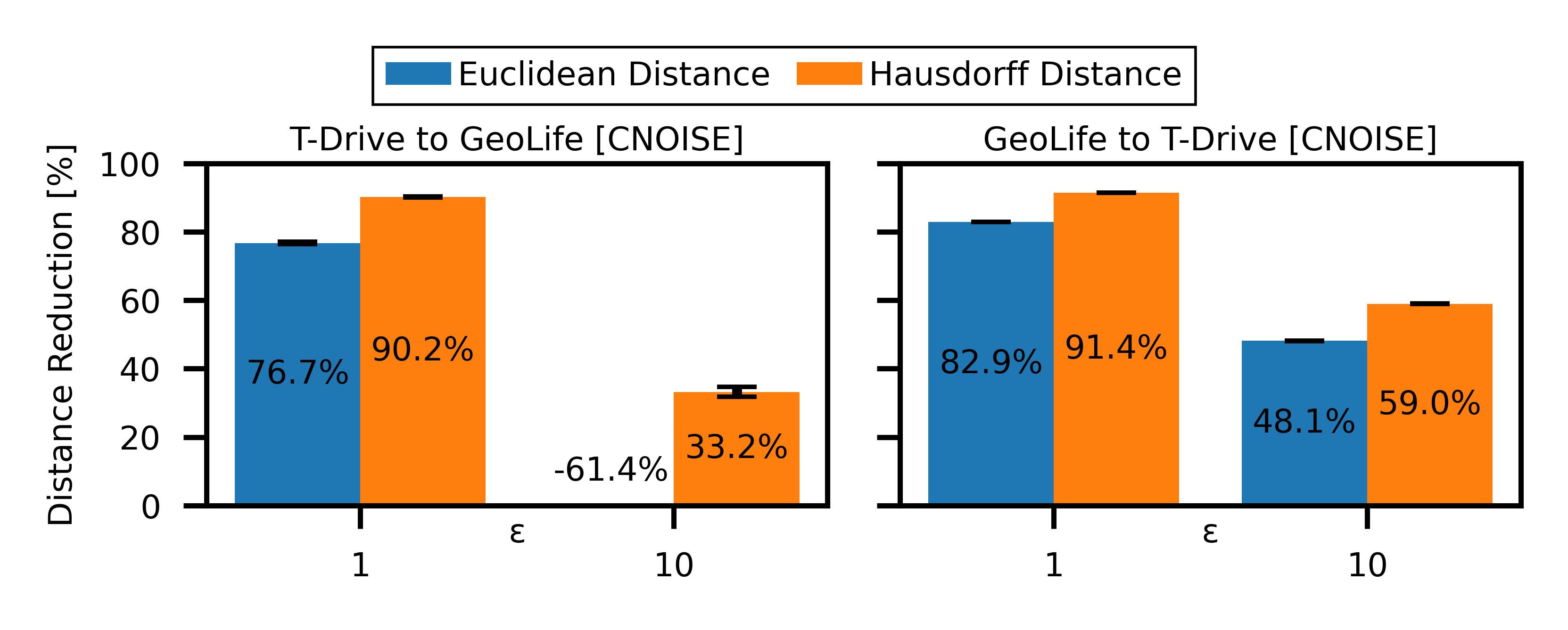}
    \caption{
    Adversary 2:
    The figure shows the results for the transfer from one dataset to another. 
    All test cases use the CNoise mechanism as protection with different choices for $\varepsilon$.
    }
	\Description[Distance reductions for adversary 2]{
	The figure shows the distance reduction of the reconstruction attack after training on one dataset and testing on the other. 
	Both test cases use the CNoise mechanism as protection with different choices for $\varepsilon$.
	For $\varepsilon = 1$, both directions achieve distance reductions of over \SI{76}{\%}.
	For $\varepsilon = 10$, the reconstruction is not successful when transferring from T-Drive to GeoLife and achieves a reduction of around \SI{50}{\%} when transferring into the opposite direction.
}
    \label{fig:a:transfer_cn}
\end{figure}

\section{Computation of Percentage Reduction}\label{sec:a:reduction-computation}

The percentage reduction stated for our evaluation measurements is computed according to the following formula:
\begin{equation*}
    \frac{(OP - OR)}{|OP|} * 100
\end{equation*}
Thereby, $OP$ refers to the \gls{OP-Distance}, and $OR$ to the \gls{OR-Distance}.
The reduction is computed for each tuple of original, protected, and reconstructed trajectory independently. Then, the average is computed over all the individual reductions.
If the percent increase of the Jaccard index is mentioned in the paper, it was computed by directly comparing the mean Jaccard index before and after reconstruction.
Computing the increase for each sample individually is not directly feasible because the Jaccard index before reconstruction might be \num{0}.

\balance %

\section{Further Results}

This section contains supplemental results for measurements not included in the main part of the paper.
First, we describe the results for adversary 2 (cf.\ Section~\ref{sec:threat-model}) and the CNoise mechanism in Section~\ref{sec:a:ad2cnosie}.
Second, we discuss our runtime measurement in Section~\ref{sec:eval:runtime}

\subsection{Dataset Transfer with CNoise Protection.}\label{sec:a:ad2cnosie}
Figure~\ref{fig:a:transfer_cn} is the corresponding figure to Figure~\ref{fig:transfer_sdd}, but with the CNoise mechanism used for protection instead of the \gls{sdd} mechanism.
As for the case with the \gls{sdd} mechanism, the results indicate that training on a different dataset lowers the reconstruction success while maintaining the general trends.
The plot also confirms that the transfer from the GeoLife dataset to the T-Drive dataset is more effective than vice-versa.
Interestingly the reconstruction from CNOISE ($\varepsilon = 10$) works with some reduction success when transferring from the GeoLife to the T-Drive dataset, but not at all in the other direction.
We do not know the cause of this result.

\subsection{Reconstruction Runtime}\label{sec:eval:runtime}
We performed our performance measurements on a single server (2x Intel Xeon Silver \num{4208}  and \SI{128}{\giga\byte} RAM) running Ubuntu 20.04.01 LTS.
The server contains \num{4} NVIDIA Tesla T4  GPUs with \SI{16}{\giga\byte} RAM each, but we only used a single GPU for the experiments.
We measured the time required for reconstruction of a single trajectory, including encoding of the protected trajectory and decoding of the reconstructed one, with a pre-trained model.
As the adversary can train the model off-line with an available dataset before the attack, the performance of training is less important than the reconstruction itself which might be performed on-line to directly execute the attack.
On the specified hardware, the runtime for the reconstruction of a single GeoLife trajectory protected with \gls{sdd} ($\varepsilon = 0.1$) lies within the \SI{99}{\%} confidence interval [\num{51.3}, \num{52.1}]  \unit{\milli\second}.
For the corresponding T-Drive trajectories (cf. Case~7), which are generally shorter, the \SI{99}{\%} confidence interval is [\num{44.8}, \num{45.6}] \unit{\milli\second}.
This short reconstruction time underlines the real-world risk of the presented reconstruction attack.

\section{All Evaluation Results}\label{sec:a:all-results}

Table~\ref{tab:all-results} displays all performed measurements.
The table specifies which dataset, protection mechanism (abbreviated with \emph{Mech.}) and $\varepsilon$ value have been used for the training and test set.
The table contains the percentage reduction of the Euclidean and Hausdorff distance after the reconstruction, which is computed as described in Appendix~\ref{sec:a:reduction-computation}.
Moreover, the table states the mean Jaccard index of the original and protected trajectories (\emph{Jaccard B.}) and of the original and reconstructed trajectories (\emph{Jaccard A.}).
A larger value for the Jaccard indicates a higher threat as discussed in Appendix~\ref{sec:a:jaccard}, and hence indicates the success of our attack.
We do not state the percentage increase for the Jaccard Index as the percentages fluctuate strongly and show very large values due to the small Jaccard indices before reconstruction.

\begin{table*}
\begin{tabular}{c|cccccccccc}
\toprule
ID  & Dataset Train & Dataset Test & Mech. Train & Mech. Test & $\varepsilon$ Train & $\varepsilon$ Test & Euclidean   & Hausdorff 
& Jaccard B. & Jaccard A.
\\ \midrule
1 & T-Drive & T-Drive & CNoise & CNoise & 0.01 & 0.01 & \SI{99.7}{\%} & \SI{99.8}{\%} & \num{1.19e-07} & \num{5.12e-02}\\
2 & T-Drive & T-Drive & CNoise & CNoise & 0.1 & 0.1 & \SI{98.1}{\%} & \SI{99.1}{\%} & \num{1.18e-05} & \num{3.44e-03}\\
3 & T-Drive & T-Drive & CNoise & CNoise & 1.0 & 1.0 & \SI{87.4}{\%} & \SI{93.4}{\%} & \num{1.17e-03} & \num{3.67e-02}\\
4 & T-Drive & T-Drive & CNoise & CNoise & 10.0 & 10.0 & \SI{65.1}{\%} & \SI{73.6}{\%} & \num{7.69e-02} & \num{2.66e-01}\\
5 & T-Drive & T-Drive & CNoise & CNoise & 100.0 & 100.0 & \SI{29.8}{\%} & \SI{29.8}{\%} & \num{5.61e-01} & \num{6.23e-01}\\
6 & T-Drive & T-Drive & SDD & SDD & 0.01 & 0.01 & \SI{68.1}{\%} & \SI{72.7}{\%} & \num{2.46e-02} & \num{7.09e-02}\\
7 & T-Drive & T-Drive & SDD & SDD & 0.1 & 0.1 & \SI{68.2}{\%} & \SI{72.8}{\%} & \num{2.46e-02} & \num{7.03e-02}\\
8 & T-Drive & T-Drive & SDD & SDD & 1.0 & 1.0 & \SI{68.4}{\%} & \SI{73.1}{\%} & \num{2.45e-02} & \num{7.13e-02}\\
9 & T-Drive & T-Drive & SDD & SDD & 10.0 & 10.0 & \SI{71.7}{\%} & \SI{77.2}{\%} & \num{2.22e-02} & \num{8.86e-02}\\
10 & GeoLife & GeoLife & CNoise & CNoise & 0.01 & 0.01 & \SI{99.4}{\%} & \SI{99.2}{\%} & \num{2.01e-10} & \num{1.72e-05}\\
11 & GeoLife & GeoLife & CNoise & CNoise & 0.1 & 0.1 & \SI{98.2}{\%} & \SI{99.1}{\%} & \num{1.96e-08} & \num{8.08e-04}\\
12 & GeoLife & GeoLife & CNoise & CNoise & 1.0 & 1.0 & \SI{91.3}{\%} & \SI{95.3}{\%} & \num{1.98e-06} & \num{1.53e-03}\\
13 & GeoLife & GeoLife & CNoise & CNoise & 10.0 & 10.0 & \SI{77.7}{\%} & \SI{82.2}{\%} & \num{1.88e-04} & \num{9.42e-03}\\
14 & GeoLife & GeoLife & CNoise & CNoise & 100.0 & 100.0 & \SI{56.5}{\%} & \SI{66.1}{\%} & \num{9.13e-03} & \num{6.78e-02}\\
15 & GeoLife & GeoLife & SDD & SDD & 0.01 & 0.01 & \SI{82.3}{\%} & \SI{87.0}{\%} & \num{3.65e-05} & \num{2.55e-03}\\
16 & GeoLife & GeoLife & SDD & SDD & 0.1 & 0.1 & \SI{83.7}{\%} & \SI{87.7}{\%} & \num{3.63e-05} & \num{2.55e-03}\\
17 & GeoLife & GeoLife & SDD & SDD & 1.0 & 1.0 & \SI{83.6}{\%} & \SI{87.7}{\%} & \num{3.60e-05} & \num{2.56e-03}\\
18 & GeoLife & GeoLife & SDD & SDD & 10.0 & 10.0 & \SI{80.2}{\%} & \SI{86.6}{\%} & \num{1.70e-05} & \num{8.86e-04}\\
19 & T-Drive & GeoLife & CNoise & CNoise & 1.0 & 1.0 & \SI{76.7}{\%} & \SI{90.2}{\%} & \num{1.98e-06} & \num{6.62e-04}\\
20 & T-Drive & GeoLife & CNoise & CNoise & 10.0 & 10.0 & \SI{-61.4}{\%} & \SI{33.2}{\%} & \num{1.88e-04} & \num{5.83e-03}\\
21 & T-Drive & GeoLife & SDD & SDD & 0.1 & 0.1 & \SI{40.1}{\%} & \SI{62.0}{\%} & \num{3.63e-05} & \num{8.01e-04}\\
22 & T-Drive & GeoLife & SDD & SDD & 1.0 & 1.0 & \SI{38.2}{\%} & \SI{61.3}{\%} & \num{3.60e-05} & \num{9.23e-04}\\
23 & GeoLife & T-Drive & CNoise & CNoise & 1.0 & 1.0 & \SI{82.9}{\%} & \SI{91.4}{\%} & \num{1.17e-03} & \num{1.66e-02}\\
24 & GeoLife & T-Drive & CNoise & CNoise & 10.0 & 10.0 & \SI{48.1}{\%} & \SI{59.0}{\%} & \num{7.69e-02} & \num{1.23e-01}\\
25 & GeoLife & T-Drive & SDD & SDD & 0.1 & 0.1 & \SI{61.0}{\%} & \SI{66.6}{\%} & \num{2.46e-02} & \num{3.18e-02}\\
26 & GeoLife & T-Drive & SDD & SDD & 1.0 & 1.0 & \SI{61.0}{\%} & \SI{66.8}{\%} & \num{2.45e-02} & \num{3.19e-02}\\
27 & T-Drive & T-Drive & CNoise & CNoise & 1.0 & 10.0 & \SI{24.3}{\%} & \SI{46.2}{\%} & \num{7.69e-02} & \num{5.45e-02}\\
28 & T-Drive & T-Drive & CNoise & CNoise & 10.0 & 1.0 & \SI{72.5}{\%} & \SI{79.3}{\%} & \num{1.17e-03} & \num{2.72e-02}\\
29 & T-Drive & T-Drive & SDD & SDD & 0.1 & 1.0 & \SI{68.4}{\%} & \SI{73.1}{\%} & \num{2.45e-02} & \num{7.21e-02}\\
30 & T-Drive & T-Drive & SDD & SDD & 1.0 & 0.1 & \SI{68.3}{\%} & \SI{72.8}{\%} & \num{2.46e-02} & \num{7.10e-02}\\
31 & T-Drive & T-Drive & CNoise & SDD & 1.0 & 1.0 & \SI{27.7}{\%} & \SI{44.9}{\%} & \num{2.45e-02} & \num{1.16e-02}\\
32 & T-Drive & T-Drive & SDD & CNoise & 1.0 & 1.0 & \SI{53.0}{\%} & \SI{70.3}{\%} & \num{1.17e-03} & \num{1.23e-02}\\
33 & T-Drive & GeoLife & CNoise & SDD & 1.0 & 0.1 & \SI{-13.8}{\%} & \SI{22.5}{\%} & \num{3.63e-05} & \num{3.92e-05}\\
34 & GeoLife & T-Drive & SDD & CNoise & 0.1 & 1.0 & \SI{49.5}{\%} & \SI{69.0}{\%} & \num{1.17e-03} & \num{1.12e-02}\\
35 & T-Drive & GeoLife & SDD & CNoise & 1.0 & 0.1 & \SI{66.0}{\%} & \SI{68.2}{\%} & \num{1.96e-08} & \num{2.86e-07}\\
36 & GeoLife & T-Drive & CNoise & SDD & 0.1 & 1.0 & \SI{22.9}{\%} & \SI{42.1}{\%} & \num{2.45e-02} & \num{9.41e-03}\\

\end{tabular}
\caption{
All evaluation cases.
This table displays all performed measurements along the percentage reduction of Euclidean and Hausdorff distance through the reconstruction,
and the mean Jaccard index [B]efore and [A]fter reconstruction.
}
\label{tab:all-results}
\end{table*}

\end{document}